\documentclass[english,aps, onecolumn]{revtex4}
\usepackage[T1]{fontenc}
\usepackage[latin9]{inputenc}
\usepackage{amsmath}
\usepackage{graphicx}
\usepackage{amssymb}

\newcommand{\tmu}{\tilde{\mu}} 
\newcommand{\tA}{\tilde{g}_A} 
\newcommand{\tB}{\tilde{g}_B} 

\newcommand{\xm}{x_{-}} 
\newcommand{\xp}{x_{+}} 

\usepackage{babel}

\begin{document}

\title{Frequency-dependent fitness induces multistability in coevolutionary
dynamics}

\author{Hinrich Arnoldt,$^{1,2}$ Marc Timme$^{1}$ and Stefan Grosskinsky$^{3}$}

\affiliation{$^{1}$Network Dynamics Group, Max Planck Institute for Dynamics
and Self-Organization (MPIDS), Bunsenstr. 10, 37073 G\"{o}ttingen, Germany}

\affiliation{$^{2}$Georg August University G\"{o}ttingen, 37077 G\"{o}ttingen, Germany}

\affiliation{$^{3}$Centre for Complexity Science, University of Warwick, Coventry
CV4 7AL, UK}
\begin{abstract}
Evolution is simultaneously driven by a number of processes such as mutation, competition and random sampling. Understanding which of these processes is dominating the collective evolutionary dynamics in dependence on system properties is a fundamental aim of theoretical research. 
Recent works quantitatively studied coevolutionary dynamics of competing species with a focus on linearly frequency-dependent interactions, derived from a game-theoretic viewpoint. However, several aspects of evolutionary dynamics, e.g. limited resources, may induce effectively nonlinear frequency dependencies. Here we study the impact of nonlinear frequency dependence on evolutionary dynamics in a model class that covers linear frequency dependence as a special case.
We focus on the simplest non-trivial setting of two genotypes and analyze the co-action of nonlinear frequency dependence with asymmetric mutation rates. We find that their co-action may induce novel metastable states as well as stochastic switching dynamics between them. Our results reveal how the different mechanisms of mutation, selection and genetic drift contribute to the dynamics and the emergence of metastable states, suggesting that multistability is a generic feature in systems with frequency-dependent fitness. 
\end{abstract}
\maketitle


\section{Introduction}

Selection, random genetic drift and mutations are the processes underlying
Darwinian evolution. For a long time population geneticists have analyzed
the dynamics in the simplest setting consisting of two genotypes evolving
under these processes \cite{Drossel2001}. In those studies, a genotype represents an
individual's genetic makeup, completely determining all relevant properties 
of the individual. A key concept
is the so-called fitness of a genotype which represents the selection
pressure for the individuals. The fitness defines the expected number of offspring
an individual will produce. Thus, selection acts on fitness
differences preferring individuals with higher fitness over individuals
with lower fitness. Usually it is assumed that individuals have fixed
fitnesses defined by their genotype alone \cite{Crow1970,Drossel2001,Blythe2007}. Yet, experimental studies
have revealed that many natural systems exhibit frequency-dependent 
selection \cite{McCauley1998,Turner1999,Gore2009}, which means that an individual's fitness
not only depends on its genotype, but also on its interactions with other individuals and hence
on the frequency of the different genotypes in the population. Although such frequency-dependent
selection had already been studied early by Crow and Kimura \cite{Crow1970}, only recently
has it received more attention \cite{Taylor2004,Traulsen2005,Antal2006,Traulsen2006,Antal2009,Antal2009_2,Assaf2010,Gokhale2011}.
In these theoretical and computational studies, individuals' interactions
are represented by interaction matrices from game theory. This
leads to a frequency dependence where the fitness depends directly on the interaction parameters in a linear way. However,
fitness may depend on many diverse factors such as cooperation (i.e. individuals acting
together to increase their fitness \cite{Traulsen2005,Traulsen2006}) and resource competition, so that
certain systems may exhibit frequency-dependent fitness that is nonlinear. 
For example, in experiments certain hermaphrodites exhibit such nonlinear fitness-dependence \cite{McCauley1998}.
To the best of our knowledge the impact of such nonlinear dependencies
on coevolutionary dynamics has not been investigated theoretically.

In this article we show that nonlinear frequency dependence \cite{Pohley1983,McCauley1998}
may induce new stable configurations of the evolutionary dynamics. Furthermore, we study
the impact of asymmetric mutation probabilities on the dynamics \cite{Sasaki2003,Durett2008},
which was also neglected in most models until now \cite{Traulsen2006,Antal2009,Antal2009_2,Gokhale2011}.
As in previous works on coevolutionary dynamics we base our work on
the Moran process in a non-spatial environment which is a well established
model to study evolutionary dynamics and was already used in many applications \cite{Nee2006}.
The Moran process is a stochastic birth-death process which
keeps the population size constant \cite{Moran1962}. 
Therefore, in a two-genotype model the system becomes effectively one-dimensional,
so that the dynamics may be described by a one-dimensional
Markov chain with transition rates defined by the Moran process. 
We derive the stationary probability distribution
of the system dynamics via its Fokker-Planck equation \cite{Risken1989}.
Sharp maxima of the distribution reveal metastable points of the dynamics
and a multitude of such maxima lead to stochastic switching dynamics
between multiple stable points.

The article is structured as follows. In Section II we introduce the
model details and in Section III we derive the Fokker-Planck equation
describing the probabilistic dynamics of the population. Using this
equation we derive the stationary probability distribution that describes
the long-time behavior of the system. In Section IV we analyze this probability distribution,
which yields information about the impact of nonlinear frequency-dependent
selection and of different mutation probabilities on the coevolutionary
dynamics. In Section V we give a summary and discuss our results.

\section{The model}

Consider a population of $N$ individuals evolving in a non-spatial
environment, meaning that the population is well-mixed so that each individual
interacts with all other individuals at all times.
In this population the individuals may assume one out
of two genotypes $A$ and $B$.
The population sizes $k_{A}$ and $k_{B}$
($k_{A}+k_{B}=N$) evolve according to the time-continuous Moran process
described in the following, cf. \cite{Moran1962}. The number of individuals $k_{A}$
of genotype $A$ completely determines the state of the system as
$k_{B}=N-k_{A}$. At all times the interactions of the individuals
determine the actual (frequency-dependent) fitness,
so that an individual's fitness of genotype $A$ or $B$ is defined by a 
fitness function $f_{A}(k_{A})$ or $f_{B}(k_{A})$ respectively.
The fitness functions $f_{A}(k_{A})$ and $f_{B}(k_{A})$ may be any
functions with the only condition that $f_{i}(k_{A})\geq0$ for all
$k_{A}\in[0,N]$ as negative fitness is not defined. At rate $f_A(k_{A})k_A$
an individual of type $A$ produces an identical offspring which may mutate
to genotype $B$ with probability $\mu_{AB}$. This applies analogously to genotype $B$.
Then one individual of the population is chosen
randomly to die, so that the population size $N$ stays constant and
the variables $k_{i}$ change maximally by $1$.
This is the so-called
Moran process which was originally introduced for a population of
two genotypes with fixed fitnesses and no mutations occurring, cf. \cite{Moran1962}.
However, this process is easily generalizable to more genotypes and frequency-dependent selection in the way
described above, cf. \cite{Taylor2004,Imhof2005}.

Note that in our model
the rate of reproduction of e.g. type $A$ is directly determined by the term
$k_A f_{A}(k_{A})$ as e.g. in \cite{Bladon2010}. In other models the fitness 
is first normalized so that the rate of reproduction is given by $k_A f_{A}(k_A)/\overline{f}(k_A)$
\cite{Traulsen2006}, where
\begin{equation}
\overline{f}(k_A)=\frac{1}{N}[k_{A}f_{A}(k_{A})+(N-k_{A})f_{B}(k_{A})] \label{eq:MeanFitness}
\end{equation}
is the population's mean fitness. While in both of these models the events occur with the same
probability, the times between the events differ by a common factor determined by the mean fitness (\ref{eq:MeanFitness}).
Thus, these models exhibit a quantitatively different time course, but the same event sequences.

Until now, usually linear functions have been considered for $f_{i}(k_{A})$.
However, in many applications nonlinear
functions seem more appropriate \cite{Pohley1983,McCauley1998}. For
example, cooperation effects in game theory induce functions linearly
increasing in $k_{A}$, so that $f_{A}(k_{A})=1+a\cdot k_{A}$ \cite{Taylor2004,Traulsen2006}.
This would mean that fitness increases infinitely with the population
size of genotype $A$. Yet, in any habitat there is only a limited
amount of resources available for living. Therefore, fitness should
decline when the population of one genotype becomes too large as
all of the individuals will compete for the same resources. We conclude
that a fitness function including these two effects has to contain
a nonlinear factor, e.g.
\begin{equation}
f_{A}(k_{A})=1+a\cdot k_{A}-b\cdot k_{A}^{2}\label{eq:fitnessfunction}\end{equation}
where $a>0$ and $b>0$. Here we choose a quadratic nonlinear factor
as a simple example. Naturally, all other nonlinear functions could
be applicable as well.

Note that linearly increasing fitness functions such as $f_{A}(k_{A})=1+a\cdot k_{A}$ and
$f_{B}(k_{A})=1+b\cdot k_{A}$ lead to quadratic relations for the \emph{mean} fitness 
$\overline{f}(k_A)$ (see equation (\ref{eq:MeanFitness})) of the population which becomes
\begin{equation}
 \overline{f}(k_{A})=1+bk_A+(a-b)\frac{k_A^2}{N}.
\end{equation}
The quadratic factor originates from the fact that the linear fitness function in (\ref{eq:MeanFitness})
is multiplied by the number of individuals leading to a quadratic dependence on $k_A$,
e.g. $k_A f_A(k_A)=k_A (1+a k_A)= k_A+a k_A^2$. This factor enters the replicator equation
in evolutionary game theory, thus leading to a quadratic fitness dependence in standard
game theoretic problems \cite{Taylor2004,Bladon2010}. However, this does not imply nonlinear
interactions and thus fitness effects such as presented in equation (\ref{eq:fitnessfunction}). Therefore, the fitness functions
that can be generated by evolutionary game theory are a special case of the functions
that occur in the theory presented here. Fitness functions as
for example presented in equation (\ref{eq:fitnessfunction}) will -- as we show in this article -- 
lead to dynamics with more complex stability properties.

Usually, when analyzing the dynamics of such a model system as described
above, it was assumed either that no mutations occur at all ($\mu_{ij}=0$)
\cite{Moran1962,Taylor2004,Assaf2010} or that the mutation probabilities
are equal ($\mu_{AB}=\mu_{BA}$) \cite{Imhof2005,Traulsen2006,Gokhale2011}.
However, usually mutation probabilities can be highly diverse \cite{Sasaki2003,Durett2008}.
This has been considered in some examples \cite{Crow1970}, but until now the effect of different 
mutation probabilities in the introduced model system has not been studied systematically.
Here, we explicitly study the effects caused by asymmetric mutation rates 
(see e.g. figure \ref{fig:AsymmetricExample}).

\section{Analysis}

In the following we assume the fitness of the individuals to be given
by
\begin{equation}
f_{A}(k)=1+g_{A}(k)\label{eq:fofk}
\end{equation}
and
\begin{equation}
 f_{B}(k)=1+g_{B}(k)
\end{equation}
with the interaction functions $g_{A}>-1$ and $g_{B}>-1$. The model
system is effectively one-dimensional with the variable $k\equiv k_{A}=N-k_{B}$
completely describing the system state \cite{Traulsen2005}. At each
event $k$ may change by at most $\pm1$ depending
only on the actual state of the system, so that the dynamics are Markovian.
Let the transition $k\rightarrow k+1$ occur with the rate $r_{k}^{+}$.
Then this rate is determined by the fitnesses and mutation probabilities
in the following way: 
\begin{itemize}
\item Each individual in the population receives offspring with rate $f_{j}(k)$
with $j\in\{A,B\}$. This means that genotype $A$ receives one offspring
with rate $k\cdot f_{A}(k)$. 
\item Such an offspring increases the population $A$ with probability $1-\mu_{AB}$, 
and population $B$ in case it mutates with probability $\mu_{AB}$.
\item One individuum is chosen uniformly to die belonging to genotype $B$ 
with probability $(N-k)/(N+1)$. 
\item Equivalently the number of individuals of genotype $A$ can increase if
genotype $B$ receives one offspring, which mutates to genotype $A$ with
probability $\mu_{BA}$, and one individuum of $B$ is chosen to die. 
\end{itemize}
Taken these processes together the transition rate for $k\rightarrow k+1$
is\begin{equation}
r_{k}^{+}=(1-\mu_{AB})kf_{A}(k)\cdot\frac{N-k}{N+1}+\mu_{BA}(N-k)f_{B}(k)\cdot\frac{N-k}{N+1}.\label{eq:plusrate}\end{equation}
Decreasing $k$ to $k-1$ is equivalent to increasing the number of
individuals of genotype $B$, so that the transition rate\begin{equation}
r_{k}^{-}=(1-\mu_{BA})(N-k)f_{B}(k)\cdot\frac{k}{N+1}+\mu_{AB}kf_{A}(k)\cdot\frac{k}{N+1}\label{eq:minusrate}\end{equation}
is obtained analogously. The master equation of the Markov
chain is then given by\begin{equation}
\frac{dp_{k}(t)}{dt}=r_{k-1}^{+}p_{k-1}(t)-(r_{k}^{-}+r_{k}^{+})p_{k}(t)+r_{k+1}^{-}p_{k+1}(t)\quad k\in\{0,N\}\label{eq:MEQ}\end{equation}
where we define $r_{-1}^{+}=r_{N+1}^{-}=0$. As also $r_{0}^{-}=r_{N}^{+}=0$,
the system has reflective boundaries at $k=0$ and $k=N$. If $\mu_{AB}=0$ ($\mu_{BA}=0$)
then $k_{A}=N$ ($k_{A}=0$) is an absorbing state of the dynamics.

It is possible to derive the exact solution to such an equation similarly to the
derivations in \cite{Claussen2005} which we do in Appendix B. However, the obtained solutions are not very explicit
and their implications are hard to grasp. It is thus more useful
to advance to the Fokker-Planck equation \cite{Risken1989} describing
the system in the limit of large $N$. As previous works have shown,
this approximation already leads to very good results for moderate
population sizes of $N<1000$, cf. \cite{Traulsen2006}. We use the transformation
\begin{eqnarray}
x=\frac{k}{N}\in[0,1],\qquad s=\frac{t}{N},\qquad\tmu_{ij}=\mu_{ij}\cdot N,\\[0.1cm]
\rho(x,s)=p_{xN}(sN)N,\qquad\tilde{g}_{j}(x)=g_{j}(xN)\cdot N \label{eq:RescaledInteraction}\end{eqnarray}
yielding the Fokker-Planck equation. The scaling factors are derived in Appendix A where the
derivation of the Fokker-Planck equation in the scaling limit
$N\rightarrow\infty$ is shown in detail (Equation (\ref{eq:AppFPE})). This leads
to\begin{equation}
\frac{d\rho(x,s)}{ds}=-\frac{\partial}{\partial x}\Big[\left(\tmu(1-2x)+\left[\tA(x)-\tB(x)\right]x(1-x)-\Delta\tmu\right)\rho(x,s)\Big]+\frac{\partial^{2}}{\partial x^{2}}\Big[x(1-x)\rho(x,s)\Big]\label{eq:FPE}\end{equation}
with the normalization condition
\begin{equation}
\int_{0}^{1}\rho(x,s)dx=1\end{equation}
for all $s\geq0$. Here $\tA=Ng_{A}(Nx)$ and $\tB=Ng_{B}(Nx)$ are the rescaled interaction
functions defined in (\ref{eq:RescaledInteraction}), $\tmu=N(\mu_{AB}+\mu_{BA})/2$ is the rescaled mean mutation
rate and $\Delta\tmu=N(\mu_{AB}-\mu_{BA})/2$ is the rescaled mutation
rate difference. These functions have to be rescaled to obtain a non-degenerate
limit, the so-called weak selection regime. 

Notice, that the drift term \begin{equation}
D^{(1)}(x)=\tmu(1-2x)+\left[\tA(x)-\tB(x)\right]x(1-x)-\Delta\tmu\end{equation}
contains three different effects. The fitness difference at each point
$x$ causes a selective drift, the mean mutation rate causes a drift
directed to $x=1/2$ and the mutation rate difference causes a one-directional
drift towards one of the boundaries $x=0$ or $x=1$. The diffusion
term\begin{equation}
D^{(2)}(x)=x(1-x)\end{equation}
reflects the undirected genetic drift.

If fitness differences are on a scale of $\mathcal{O}\left(N^{-1}\right)$, then
mutation, selection and genetic drift all act on the same scale
\cite{Ohta2002,Traulsen2006}. Then the dynamics are characterized by an interplay 
of these different effects.
On the other hand, in the strong selection
regime -- where fitness differences are on a scale $\mathcal{O}(1)$
-- genetic drift becomes negligible and the dynamics in the bulk of
the system become deterministic for large $N$ \cite{Jain2007}. However, as the factor 
$x(1-x)$ vanishes for $x\rightarrow0$ and $x\rightarrow1$ the effects of mutations play an important role. 
Clearly, without mutations $x=0$ and $x=1$ are absorbing states, but even in the strong selection
regime they become non-absorbing for any positive mutation rate. This is reflected in the form of the drift 
term $D^{(1)}(x)$, as only the factors reflecting mutations $\tmu(1-2x)-\Delta\tmu$ remain non-zero in
the limits $x\rightarrow0$ and $x\rightarrow1$.

For a one-dimensional Fokker-Planck equation (\ref{eq:FPE}) the stationary distribution 
is \cite{Risken1989}\begin{equation}
\rho^{\ast}(x)=Ce^{-\Phi(x)}\end{equation}
with the potential\begin{equation}
\Phi(x)=\ln\left(D^{(2)}(x)\right)-\int\frac{D^{(1)}(x)}{D^{(2)}(x)}dx\end{equation}
which is defined up to a constant irrelevant to our calculations as
it is canceled by the normalization\begin{equation}
C=\frac{1}{\int_{0}^{1}e^{-\Phi(x)}dx}.\label{eq:Normalization}\end{equation}
Thus, we obtain\begin{eqnarray*}
\Phi(x) & = & \ln\left(x(1-x)\right)-\int\left[\frac{\tmu(1-2x)}{x(1-x)}+\tA(x)-\tB(x)-\frac{\Delta\tmu}{x(1-x)}\right]dx\\
 & = & \ln\left(x(1-x)\right)-\tmu\ln(x(1-x))+\Delta\tmu\ln\left[\frac{x}{1-x}\right]-\int\left[\tA(x)-\tB(x)\right]dx\end{eqnarray*}
where we used\begin{equation}
(1-2x)=\frac{\partial}{\partial x}[x(1-x)].\end{equation}
 The stationary solution becomes\begin{equation}
\rho^{\ast}(x)=C\left(x(1-x)\right)^{\tmu-1}\left(\frac{x}{1-x}\right)^{-\Delta\tmu}e^{\int\left[\tA(x)-\tB(x)\right]dx}\label{eq:GeneralSolution}\end{equation}
where the constant $C$ defined in (\ref{eq:Normalization}) has to
be calculated numerically for given parameters.

This stationary distribution contains contributions from the above mentioned four different
effects:
\begin{enumerate}
\item Genetic drift is reflected by $(x(1-x))^{-1}$ which diverges for $x\in\{0,1\}$ at the boundaries.
\item The mean mutation rate $\tmu>0$ causes the balancing term $(x(1-x))^{\tmu}$.
\item The asymmetry in the mutation probabilities causes the term $(x/(1-x))^{-\Delta\tmu}$ which diverges at one boundary and vanishes at the other.
\item All frequency-dependent selection effects are contained in the exponential
factor $e^{\int\left[\tA(x)-\tB(x)\right]dx}$ which can take various shapes.
\end{enumerate}

\section{Multiple stable points}

What are the possible shapes for the stationary distribution $\rho^{\ast}(x)$
in the form given by equation (\ref{eq:GeneralSolution})? Until now,
the term $(x/(1-x))^{-\Delta\tmu}$ representing the asymmetric mutation
probabilities has not been considered to our knowledge and the interaction
functions $\tA(x)$ and $\tB(x)$ in the selection term were considered
to be at most linear in $x$ \cite{Traulsen2006}. We should therefore
be interested in the effects of nonlinear interaction functions and
asymmetric mutation rates.

\begin{figure}
\includegraphics[width=15cm]{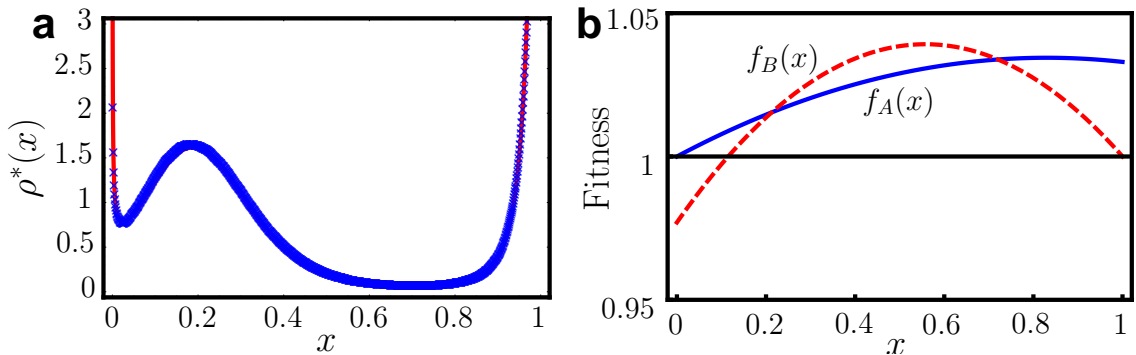}
\caption{The stationary distribution for the model system exhibits 3 maxima corresponding to metastable
points. (a) shows the theoretical curve from equation (\ref{eq:GeneralSolution})
(red, solid) in perfect agreement with data from simulations with
$N=1000$ (blue, $\times$). (b) shows the fitness functions of genotype
$A$ (blue, solid) and $B$ (red, dashed). Interaction parameters are $a_{A}=0.083$,
$b_{A}=0.05$, $a_{B}=0.177$ and $b_{B}=0.2$ (see equation (\ref{eq:fofk}) and (\ref{eq:Interactions}))
while the mutation rate is $\tmu=0.5$ ($\Delta\tmu=0$).}
\label{fig:FirstExample}
\end{figure}

Let us first analyze the dynamics for nonlinear interaction functions
\cite{Pohley1983,McCauley1998} describing the effects of cooperation
and limited resources as described by equation (\ref{eq:fitnessfunction}),
so that\begin{equation}
\tA(x)=N\left(a_{A}x-b_{A}x^{2}\right)\qquad\tB(x)=N\left(a_{B}(1-x)-b_{B}(1-x)^{2}\right).\label{eq:Interactions}\end{equation}
Already with $\Delta\tmu=0$ such interaction functions induce dynamics
stochastically switching between three metastable points. An example
is shown in Figure \ref{fig:FirstExample}, where the theoretically calculated
stationary distribution from equation (\ref{eq:GeneralSolution})
is shown together with data from simulations with a population of
$N=1000$ individuals which is enough to obtain almost perfect fitting 
(cf. also Figure \ref{fig:NPlot}).
The fitness functions (Figure \ref{fig:FirstExample}b) both show at first
an increase on increasing the number of individuals of genotype $A$
or $B$ from $0$ due to cooperation effects and then a strong decrease
due to resource competion. As the resulting fitness functions are
asymmetric, also the stationary distribution is asymmetric. There is a
maximum at $x=0$ due to genetic drift. Selection drives the dynamics towards a metastable state
at $x\approx0.2$, because for $x<0.2$ genotype $A$ is fitter than $B$ thus increasing in frequency
and for $x>0.2$ genotype $A$ is less fit than $B$ thus decreasing in frequency (cf. Figure \ref{fig:FirstExample}b). 
The maximum of the stationary distribution is thus exactly at the point where $f_A(x)=f_B(x)$. For $x>0.7$
again genotype $A$ is fitter than $B$ and thus the dynamics are driven 
towards $x=1$ by selection as well as genetic drift. The mutational
force induced by $\tmu=0.5$ increases the height of the maximum at
$x\approx0.2$ driving the system away from the maxima at $x=0$ and
$x=1$. So, in this example genetic drift, mutation and selection
all significantly influence the dynamics.

\begin{figure}
\includegraphics[width=15cm]{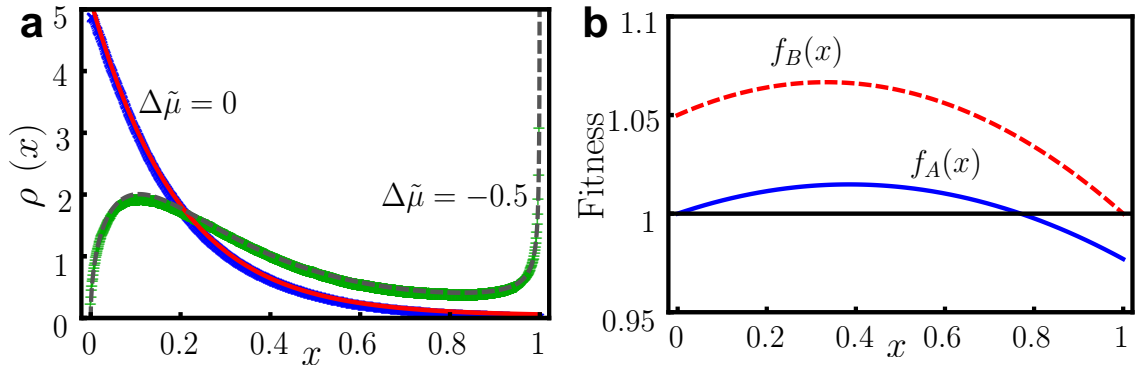}
\caption{Asymmetric mutation rates cause the emergence of a new maximum of
the stationary distribution. (a) shows the stationary distribution
of the Fokker-Planck equation for $\Delta\tmu=0$ (red, solid) and $\Delta\tmu=-0.5$
(gray, dashed) for a system with selective advantage for genotype $B$ as
shown in (b) (blue, solid: genotype $A$; red, dashed: genotype $B$). Solid and dashed lines
in (a) show the theoretical curves from equation (\ref{eq:GeneralSolution}),
crosses the data from simulations with $N=1000$ ($\times$: $\Delta\tmu=0$, $+$: $\Delta\tmu=-0.5$).
Interaction parameters are $a_{A}=0.077$, $b_{A}=0.1$, $a_{B}=0.2$
and $b_{B}=0.15$ (see equation (\ref{eq:Interactions})) while mean
mutation rate is $\tmu=1$.}
\label{fig:AsymmetricExample}
\end{figure}

Let us now study the influence of the mutation rates in detail. Interestingly,
for asymmetric mutation rates ($\Delta\tmu\neq0$) the factor $(x/(1-x))^{-\Delta\tmu}$
always diverges in the interval $[0,1]$. For $\Delta\tmu>0$ it diverges
for $x\rightarrow0$, otherwise for $x\rightarrow1$. This can cause
the emergence of a maximum of the stationary distribution at $x=0$
(or resp. $x=1$). Figure \ref{fig:AsymmetricExample} shows an example where
due to asymmetric mutation rates a maximum occurs at $x=1$, when
for $\Delta\tmu=0$ there is an absolute minimum at $x=1$ as
this stable state minimizes the population's mean fitness. This
means, that the system dynamics are mutation dominated near $x=1$.
Furthermore, the maximum located at $x=0$ for $\Delta\tmu=0$ is
shifted for $\Delta\tmu>0$ to values $x>0$ causing a minimum of
the stationary distribution at $x=0$. Thus, in this example both
selection and the asymmetry in mutation probabilities are the driving
forces of the system's dynamics.

\begin{figure}
\includegraphics[width=15cm]{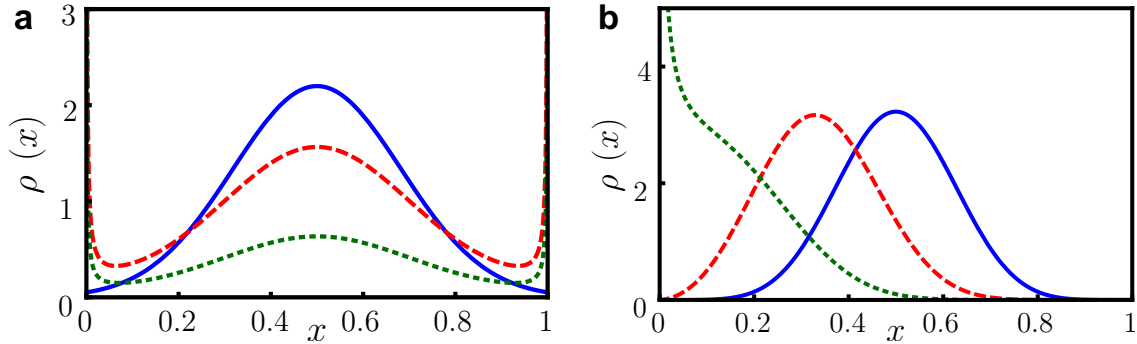}
\caption{The mutation rates strongly influence the system's dynamics. (a) shows
stationary solutions of an example system with interaction functions as in equation (\ref{eq:Interactions})
for three different symmetric ($\Delta\tmu=0$) mutation rates: $\tmu=1$ (blue, solid), $\tmu=0.1$ (red, long dashed)
and $\tmu=0.01$ (green, short dashed). This illustrates that for low mutation rates the dynamics stay close
to the system's edges for long times. Here, differences in the mutation rates only slightly affect
the shape of the stationary distribution. On the other hand, (b) shows that for high mutation rates ($\tmu=5$)
the dynamics are drawn stronger towards the middle. Here, differences in the mutation rates have a 
stronger impact, as the three curves with $\Delta\tmu=0$ (blue, solid), $\Delta\tmu=2.5$ 
(red, long dashed) and $\Delta\tmu=4.5$ (green, short dashed)
demonstrate. The increasing $\Delta\tmu$ shifts the existing stable state towards the edge of the system
until it vanishes. System parameters were $N=1000$ and a symmetric interaction function according to
equation (\ref{eq:Interactions}) with one stable state at $x=0.5$ with $a_A=a_B=-0.01$, $b_A=b_B=0.005$.}
\label{fig:MutPlot}
\end{figure}

\begin{figure}
\includegraphics[width=15cm]{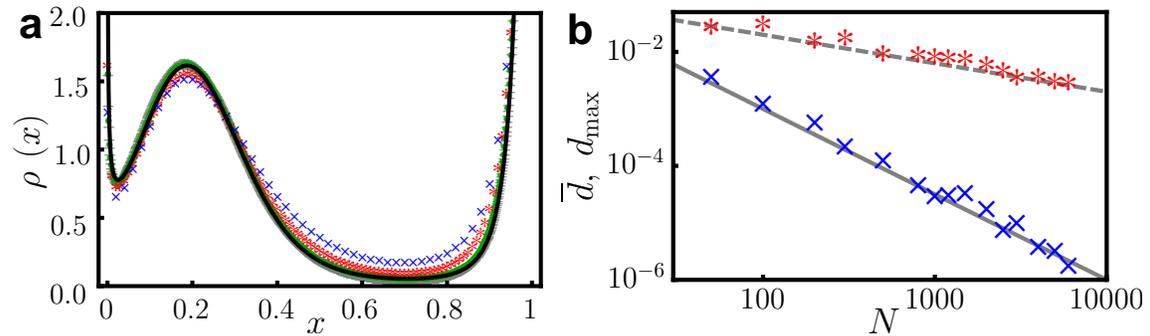}
\caption{The theoretically obtained stationary distribution well fits data from simulations for large population sizes $N$.
(a) shows the theoretical curve from equation (\ref{eq:GeneralSolution})
(black, solid) for the same system as in Figure \ref{fig:FirstExample}. Data points show empirical distributions as defined
in equation (\ref{eq:empiricaldensity}) obtained in simulations for $N=50$ (blue, $\times$), $N=100$ (red, $\ast$),
$N=500$ (green, $\bullet$) and $N=1000$ (gray, $+$).
(b) shows the distances $\overline{d}$ and $d_{\text{max}}$ of simulation data and theoretical curve for different $N$ 
for the mean distance measure $\overline{d}$ defined in equation (\ref{eq:meandistance}) (blue, $\times$) and the maximum distance measure
$d_{\text{max}}$ defined in equation (\ref{eq:maxdistance}) (red, $\ast$), demonstrating that the distance decays for increasing $N$. 
The solid line $N^{-3/2}$ and the dashed line $2N^{-1/2}$ are added as a guide to the eye for the relation between measured distances 
and system size $N$. The measured empirical distributions for both (a) and (b) were obtained by simulating the system dynamics 
from an initial state drawn from $\rho^\ast(x)$ for a mixing time $T_{\text{mix}}=100N$ and then recording the density for a 
time $T_{\text{meas}}=10N^2$.}
\label{fig:NPlot}
\end{figure}

For low mutation rates the population has a high tendency to become dominated by one 
of the two genotypes for long times which is illustrated in Figure \ref{fig:MutPlot}a.
Thus, the dynamics stay close to the edges $x=0$ or $x=1$ waiting for one of
the few mutations to occur. For low overall mutation rates differences in the 
mutation rates have only weak effects. High mutation rates
cause the population to be drawn towards a mixture of both genotypes ($x\approx0.5$).
Even more, for high mutation rates the differences in the mutation rates $\Delta \tmu$ can 
have an important influence on the dynamics. For instance Figure \ref{fig:MutPlot}b illustrates
a shift of an existing stable point, which completely vanishes for
high $\Delta\tmu$. We conclude that asymmetric mutation
rates can cause the emergence of new stable states (as in Figure \ref{fig:AsymmetricExample}),
the disappearance of existing ones (as in Figure \ref{fig:MutPlot}), 
and shift existing stable states to new positions.

As we saw above, the fit of the theoretically calculated stationary distribution
to simulation data is almost perfect for weak selection, i.e. the maximal fitness difference 
satisfies $\Delta f_{\text{max}}\ll1$, cf. e.g. \cite{Traulsen2006}. To quantify the quality of the fit we 
measured the stationary distribution with the same parameters as above for different
population sizes $N$. This is shown in Figure \ref{fig:NPlot} demonstrating that
for increasing $N$ the data gets ever closer to the theoretically obtained 
distribution. For this we define the empirical distribution
\begin{equation}
 \pi_k:=\frac{1}{T_{\text{meas}}}\sum_{t=0}^{T_{\text{meas}}}\delta(X_{t+T_{\text{mix}}},k)\label{eq:empiricaldensity}
\end{equation}
where $(X_t:\,t\geq0)$ is the discrete process defined in equation (\ref{eq:MEQ}). $T_{\text{mix}}$
is a time large enough for the process to reach stationarity and depends on the system size, as does the measurement time $T_{\text{meas}}$ which is specified in Figure \ref{fig:NPlot}. Using this definition we quantify the quality of the fit using the distance measure
\begin{equation}
 \overline{d}:=\frac{1}{N}\sum_{k=1}^{N-1}\left|\pi_k-\int_{k/N-1/2N}^{k/N+1/2N}\rho^\ast(x)dx\right| \label{eq:meandistance}
\end{equation}
which takes the mean distance of the measured empirical distribution $\pi_k$ from the
theoretical distribution for every point except $k=0$ and $k=N$ where the density $\rho^\ast(x)$ diverges. Here, 
the the theoretical distribution is calculated by integrating the theoretical density over bins of the size 
$1/N$ around the points $k/N$. Figure \ref{fig:NPlot}b shows that the mean distance $\overline{d}$ decays with increasing $N$.

The decay of this quantity is dominated by the slow convergence at the domain boundaries $k=0$ and $k=N$. Therefore,
we additionally defined the maximum distance measure
\begin{equation}
 d_{\text{max}}:=\max_{k\in[1,N-1]}\left\{\left|\pi_k-\int_{k/N-1/2N}^{k/N+1/2N}\rho^\ast(x)dx\right|\right\}\label{eq:maxdistance}
\end{equation}
which gives the maximum distance between the measured probability distribution from 
the theoretical distribution for all points except $k=0$ and $k=N$ due to the same argument as above.
Figure \ref{fig:NPlot}b shows that the maximum distance $d_{\text{max}}$ 
decreases with increasing $N$, however slower than the mean distance $\overline{d}$ due to the divergence of $\rho(x)$ 
at the boundaries. Altogether we conclude, that the theoretical curve fits the data very well already for
$N\succsim1000$. However, near the domain boundaries special care has to be taken, as divergences of the theoretical
curve lead to larger deviations.

Furthermore, as the Fokker-Planck approximation only holds for weak selection we study
the quality of the solution obtained throught the Fokker-Planck equation in dependence of the selection strength. For this we
introduce a scaling factor $\xi$ to the interaction functions, so that they become
\begin{equation}
 \tA(x)=\xi N\left(a_{A}x-b_{A}x^{2}\right)\qquad\tB(x)=\xi N\left(a_{B}(1-x)-b_{B}(1-x)^{2}\right)\label{eq:ScaledInteractions}.
\end{equation}
We find that the stationary solution of the Fokker-Planck equation well approximates the solution
of the Master equation
\begin{equation}
 p_k^\ast=\frac{\prod_{j=0}^{k-1}\frac{r_j^+}{r_{j+1}^-}}{\sum_{l=0}^N \prod_{j=0}^{l-1}\frac{r_j^+}{r_{j+1}^-}}\label{eq:MasterSolutionInText}
\end{equation}
for selection strengths up to $\xi$ of the order of 1 (cf. Figure \ref{fig:SelectionPlot} ).
We derive the solution (\ref{eq:MasterSolutionInText}) of the Master equation
in Appendix B. Note that for the derivation of this solution no approximation is necessary and it hence fits data
from simulations perfectly up to an error due to finite sampling in simulations. 
As Figure \ref{fig:SelectionPlot} illustrates, for strong selection $\xi\succsim1$ the solution of the Fokker-Planck equation does not fit the exact
solution of the Master equation perfectly, but still catches the overall trend of the dynamics.
The error, quantified by the mean distance measure analogous to (\ref{eq:meandistance}), 
increases with $\xi$ as a power law both in comparison with
the solution from the Master equation and with data from simulations, cf. Figure \ref{fig:SelectionPlot}d. 
All in all, for weak selection the Fokker-Planck approximation works well
while for strong selection it does not perfectly predict the stationary distribution, but still reflects the overall
trend of the dynamics. If this approximation is not satisfying, the direct solution (\ref{eq:MasterSolutionInText})
of the Master equation yields the exact distribution fitting the data for all selection strengths.

\begin{figure}
\includegraphics[width=15cm]{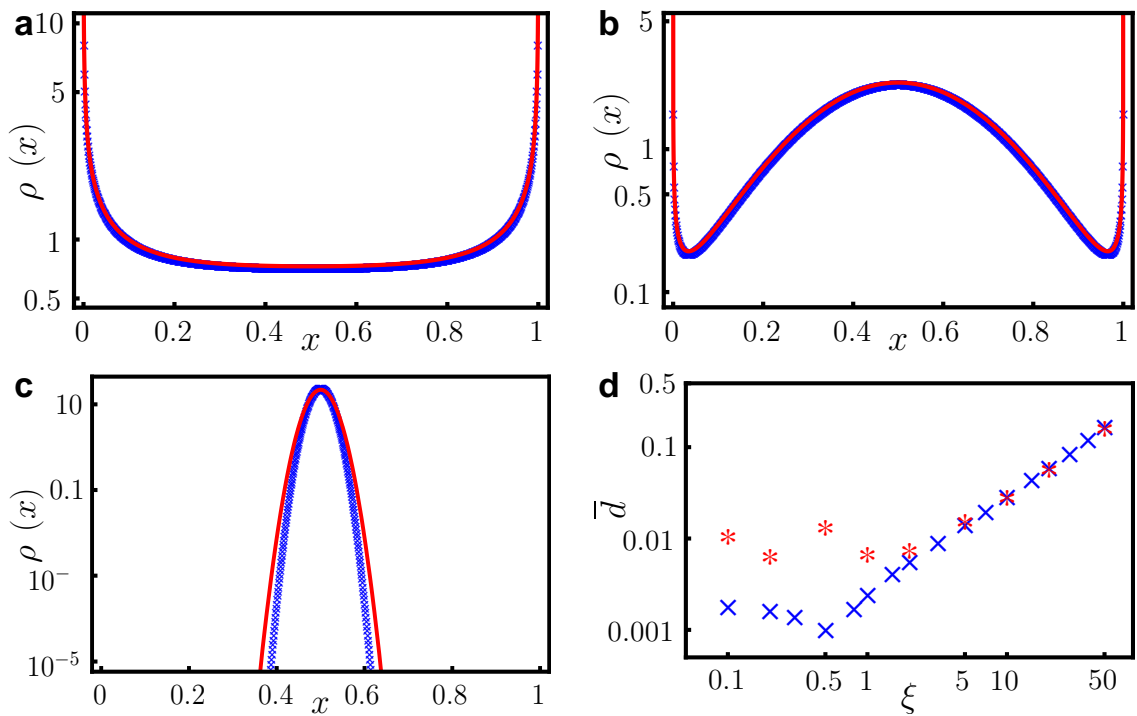}
\caption{The stationary distribution (\ref{eq:GeneralSolution}) obtained from the Fokker-Planck equation well fits 
the exact solution (\ref{eq:MasterSolutionInText}) from the Master equation for weak selection, but
not for strong selection. (a) shows the stationary solution from the Fokker-Planck equation (red, solid) 
for the system with interaction functions defined by equation (\ref{eq:ScaledInteractions}) for very weak selection 
$\xi=0.1$ together with the solution from the Master equation (blue, $\times$). The distributions of the same system for weak selection
$\xi=1$ and strong selection $\xi=50$ are shown in (b) and (c), respectively. All curves are plotted logarithmically
to better allow a comparison of the deviations over all scales. (c) shows that the curve from the Fokker-Planck equation does not
fit the exact Master solution very well. This is quantified in (d) showing the mean distance $\overline{d}$ between Fokker-Planck
and Master solution (blue, $\times$) and between Fokker-Planck solution and an empirical density obtained from simulations 
(red, $\ast$). The system parameters were $N=1000$, $\tmu=0.5$, $a_A=a_B=-0.01$ and $b_A=b_B=0.005$.
 The measured empirical distributions were obtained by simulating the system dynamics from an initial state drawn from 
$\rho^\ast(x)$ for a mixing time $T_{\text{mix}}=10^5$ and then recording the density for a time $T_{\text{meas}}=10^7$.}
\label{fig:SelectionPlot}
\end{figure}

\begin{figure}
\includegraphics[width=15cm]{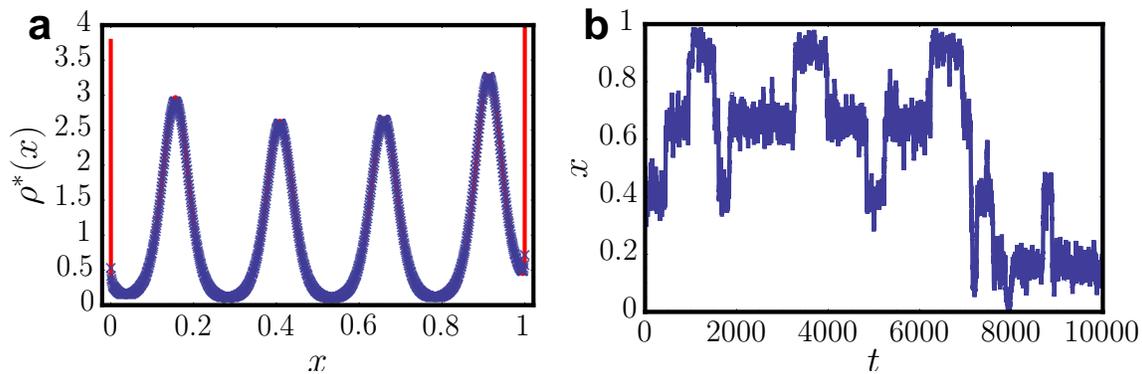}

\caption{Multiple stability dynamics for periodic interaction functions. (a)
shows the theoretically computed stationary distribution from equation
(\ref{eq:GeneralSolution}) (red, solid) together with simulation
data with $N=1000$ (blue, $\times$) for the interaction functions given
in (\ref{eq:PerFitFunc}). The computed stationary distribution diverges for $x\rightarrow0$ and $x\rightarrow1$ while the simulation data remains finite due to the finite number of individuals. (b) shows a sample path which exhibits switching
between the different maxima of the distribution. Parameters are $\alpha=30$,
$\beta=25$, $\tmu=0.8$ and $\Delta\tmu=0$.}
\label{fig:MultipleStability}
\end{figure}

We have shown above that the dynamics
of a two-genotype system can exhibit multiple stable points
induced by nonlinear selection. Even more, there is no theoretical limit to the number of stable points
in the system. Assume for example $\tmu=1$ and $\Delta\tmu=0$, so that mutation exactly balances genetic drift. 
Then, the stationary distribution (\ref{eq:GeneralSolution}) has a maximum
at each point, where $\exp\left[\int\tA(x)-\tB(x)dx\right]$ has a
maximum. Theoretically $\tA(x)$ and $\tB(x)$ can be any function
with an arbitrary amount of extreme points in $[0,1]$, so that there
is no limit to the stationary distribution's number of maxima, if
selection dominates the dynamics. However, for finite $N$ the number of possible maxima is naturally limited by $N/2$.
Figure \ref{fig:MultipleStability}
shows an example, where we used periodic interaction functions\begin{equation}
\tA(x)=\alpha\left(1+\sin(\beta x)\right)\text{ and }\tB(x)=\alpha\left(1+\cos(\beta x)\right).\label{eq:PerFitFunc}\end{equation}
 Although this is not a realistic interaction function in most applications,
it demonstrates what is theoretically possible in the introduced system.

\section{Summary}
In this article we analyzed a two-genotype system in a very general setting with
(possibly) asymmetric mutation probabilities and nonlinear fitness
functions \cite{Pohley1983,McCauley1998} in finite populations. 
The underlying Moran process is a well established model \cite{Drossel2001,Traulsen2005,Traulsen2006}
to gain an understanding of the interplay of selection, mutation and genetic drift
in evolutionary dynamics. However, the Moran process is studied mostly with
symmetric mutation probabilities and at most linear interaction functions.
We reasoned that neither need mutation probabilities be symmetric -- as 
experiments have shown, that mutation probabilities are often asymmetric \cite{Durett2008} --
nor can all interaction effects be described by linear interaction functions.
For example cooperation in game theory leads to an interaction function
increasing linearly in the frequency of the cooperating genotype.
Yet, in many applications also cooperators in the end compete for
the same type of resource which is limited. Therefore, due to limited
resources a population being too large cannot be sustained leading
to a decrease in the fitness. There is no linear function that can
reflect both of these effects at the same time.

We derived the Fokker-Planck equation describing the dynamics of the
number of individuals $k$ of genotype $A$ in the limit of large
population sizes $N$. We quantified the quality of the Fokker-Planck approach
for an example (cf. Figure 4) where the difference of simulation data and
theoretical solution became almost not detectable for population sizes larger than
$N\succsim1000$. Actually, if the system exhibits absorbing states then the Fokker-Planck method
does not work to study the corresponding quasi-stationary distributions.
Instead, WKB methods are more appropriate to describe the system dynamics
as for example in \cite{Assaf2010}, where fixation resulting from large fluctuations
was studied. In our model system no such absorbing states exist, as long as the mutation rates
are positive ($\mu_{ij}\neq0$) and therefore the Fokker-Planck equation is 
appropriate to describe the system dynamics.

We identified the individual effects of selection, mean mutation rate and mutation
difference as well as genetic drift and derived the stationary probability distribution
as determined by the Fokker-Planck equation.
Analyzing the distribution, we found that asymmetries in the mutation
probabilities may not only induce the shifting of existing stable
points of the dynamics to new positions, but also lead to the emergence
of new stable points. Thus, a genotype that has a selective disadvantage
can anyway have a stable dynamical state where its individuals dominate
the population due to a higher mutational stability (see Figure \ref{fig:AsymmetricExample}). Further we found, that
dynamic fitness leads to multiple stable points of the dynamics induced
by selection and also genetic drift. We showed an example (Figure
\ref{fig:FirstExample}) where three stable points exist, two caused by
selection and one by genetic drift.

We conclude that frequency-dependent fitness together with asymmetric
mutation rates induces complex evolutionary dynamics, in particular if the
interactions imply nonlinear fitness functions. Theoretically, there is no limit
for the number of stable points that the dynamics can exhibit (see
Figure \ref{fig:MultipleStability}). 
All in all, we interprete our results such that in real biological systems multiple metastable equilibria may exist,
whenever species interact in a way complex enough to imply a fitness that nonlinearly depends on frequency.
As a consequence, one species may exhibit a certain frequency for a long time 
before a sudden shift occurs and then a new frequency prevails. Such a change may thus occur even in the absence of
changes of the environment; it may be induced as well by a stochastic switching from one metastable
state to another due to complex inter-species interactions.

The Moran process is a standard tool
to gain theoretical insights into experimental data \cite{Nee2006}. Of course, we do not propose here that any experimental setup
may be exactly described by the Moran process. However, we think that it should be feasible to develop an experiment
where two different mutants evolve with asymmetric mutation rates. To find an experimental
setup where the two genotypes also exhibit nonlinear fitness could however prove more difficult. 
Rather, our study is a theoretical study indicating that nonlinear fitness
may be the cause of multible stable states when observed in experimental data.

For further studies on systems
with more genotypes it should be useful to combine our considerations presented here with the work
of Traulsen et al. \cite{Traulsen2006}, where an analysis of systems
with more than two genotypes was carried out. Extending those results it
may be possible to gain a better understanding of the effects of
nonlinear interactions for many different genotypes. Also, it may
be interesting to study the effects of changing interactions, where
the interactions change according to the system dynamics \cite{Gross2008}.
Thus, our study might serve as a promising starting point to investigate
how nonlinear frequency dependencies impact evolutionary dynamics in complex environments.

\textbf{Acknowledgements}

We thank Steven Strogatz for fruitful discussions during project initiation and Stefan Eule for helpful technical discussions.
Stefan Grosskinsky acknowledges support by EPSRC, grant no. EP/E501311/1.

\section{Appendix A -- The Fokker-Planck Equation}

The master equation (\ref{eq:MEQ}) may be transformed to a Fokker-Planck
equation in the limit of large $N$ \cite{Risken1989}. We introduce
the transformation\begin{equation}
x=\frac{k}{N}\qquad s=t\cdot F(N)\qquad\tmu_{ij}=\mu_{ij}\cdot G(N)\end{equation}
together with the rescaled functions\begin{equation}
\rho(x,s)=p_{Nx}(t)N\qquad\tilde{g}_{j}(x)=g_{j}(k)\cdot H(N).\end{equation}
We fix the scaling functions $F(N)$, $G(N)$ and $H(N)$ such that
in the limit $N\rightarrow\infty$ all terms in equation (\ref{eq:MEQ})
remain finite so that mutation, selection and genetic drift all act
on the same scale. We further define\begin{equation}
x_{-}=x-\frac{1}{N}\qquad x_{+}=1+\frac{1}{N}\end{equation}
and substituting all this into the master equation (\ref{eq:MEQ}),
we obtain\begin{eqnarray}
\frac{d\rho(x,s)}{ds}F(N) & = & \frac{N^{2}}{N+1}\left\{ \left[(1-\mu_{AB})(1+g_{A}(\xm))\xm(1-\xm)+\mu_{BA}(1+g_{B}(\xm))(1-\xm)^{2}\right]\rho(\xm,s)\right.\nonumber \\
 &  & \left[(1-\mu_{BA})(1+g_{B}(\xp))(1-\xp)\xp+\mu_{AB}(1+g_{A}(\xp))\xp^{2}\right]\rho(\xp,s)\nonumber \\
 &  & -\left[(1-\mu_{AB})(1+g_{A}(x))x(1-x)+\mu_{BA}(1+g_{B}(x))(1-x)^{2}\right.\nonumber \\
 &  & +\left.\left.(1-\mu_{BA})(1+g_{B}(x))x(1-x)+\mu_{AB}(1+g_{A}(x))x^{2}\right]\rho(x,s)\right\} \label{eq:App1}\end{eqnarray}
We choose $F(N)=1/(N+1)$ and $G(N)=H(N)=N$ so that in the limit
$N\rightarrow\infty$ the terms stay finite. Further we introduce
the mean mutation rate $\tmu=N(\mu_{AB}+\mu_{BA})/2$ and the mutation
rate difference $\Delta\tmu=N(\mu_{AB}-\mu_{BA})/2$. To not overload
the notation we drop the time argument $s$ of $\rho(x,s)$ in the
following calculation. This leads to \begin{eqnarray*}
\frac{d\rho(x)}{ds} & = & N^{2}\left\{ -2x(1-x)\rho(x)+\xp(1-\xp)\rho(\xp)+\xm(1-\xm)\rho(\xm)\right.\\
 &  & +\left.\frac{\tmu}{N}\left[-(1-2x)^{2}\rho(x)+\frac{1}{2}(1-2\xp)^{2}\rho(\xp)+\frac{1}{2}(1-2\xm)^{2}\rho(\xm)\right]\right\} \\
 &  & +N\left\{ \tA(\xm)\xm(1-\xm)\rho(\xm)-\tA(x)x(1-x)\rho(x)+\tB(\xp)\xp(1-\xp)\rho(\xp)-\tB(x)x(1-x)\rho(x)\right.\\
 &  & +\frac{\tmu}{2}\left[(1-2\xm)\rho(\xm)-(1-2\xp)\rho(\xp)\right]\\
 &  & +\Delta\tmu\left[\xp\rho(\xp)-x\rho(x)-(1-\xm)\rho(\xm)+(1-x)\rho(x)\right]\\
 &  & +\frac{\tmu}{N}\left[\left(\tA(\xp)\xp^{2}+\tB(\xp)(\xp^{2}-\xp)\right)\rho(\xp)-\left(\tA(x)x^{2}+\tB(x)(x^{2}-x)\right)\rho(x)\right.\\
 &  & +\left.\left(\tA(\xm)(\xm^{2}-\xm)+\tB(\xm)(1-\xm)^{2}\right)\rho(\xm)-\left(\tA(x)(x^{2}-x)+\tB(x)(1-x)^{2}\right)\rho(x)\right]\\
 &  & +\frac{\Delta\tmu}{N}\left[\left(\tA(\xp)\xp^{2}-\tB(\xp)(\xp^{2}-\xp)\right)\rho(\xp)-\left(\tA(x)x^{2}-\tB(x)(x^{2}-x)\right)\rho(x)\right.\\
 &  & +\left.\left.\left(\tA(\xm)(\xm^{2}-\xm)-\tB(\xm)(1-\xm)^{2}\right)\rho(\xm)-\left(\tA(x)(x^{2}-x)-\tB(x)(1-x)^{2}\right)\rho(x)\right]\right\} \end{eqnarray*}
 In the limit $N\rightarrow\infty$ the different terms with $N^{2}$
in front become second order derivatives with respect to $x$, while
the other terms become first order derivatives. The terms which have
a $1/N$ factor vanish in the limit $N\rightarrow\infty$ and thus
the above equation becomes \begin{equation}
\frac{d\rho}{ds}=-\frac{\partial}{\partial x}\left[\tmu(1-2x)\rho(x)+\left(\left[\tA(x)-\tB(x)\right]x(1-x)\rho(x)\right)-\Delta\tmu\rho(x)\right]+\frac{\partial^{2}}{\partial x^{2}}\left[x(1-x)\rho(x)\right]\label{eq:AppFPE}\end{equation}
which is the Focker-Planck-Equation of the system.

\section{Appendix B -- Stationary Solution of the Master Equation}
We directly derive the stationary solution $p_k^\ast$ of the master equation (\ref{eq:MEQ})
using the detailed balance equation
\begin{equation}
 r_{k-1}^+ p_{k-1}^\ast=r_{k}^- p_{k}^\ast
\end{equation}
which applies to any chain with only nearest neighbour transitions \cite{Risken1989}, cf. also \cite{Claussen2005}.
Thus, using the rewritten balance equation
\begin{equation}
 p_{k}^\ast=\frac{r_{k-1}^+}{r_{k}^-}\cdot p_{k-1}^\ast
\end{equation}
iteratively, we obtain
\begin{equation}
 p_k^\ast=p_0^\ast\prod_{j=0}^{k-1}\frac{r_j^+}{r_{j+1}^-}.
\end{equation}
Finally, we may use the normalization condition
\begin{equation}
 \sum_{l=0}^N p_l^\ast=1
\end{equation}
of the stationary distribution to eliminate the factor $p_0^\ast$. We then obtain the
exact stationary solution of the master equation
\begin{equation}
 p_k^\ast=\frac{\prod_{j=0}^{k-1}\frac{r_j^+}{r_{j+1}^-}}{\sum_{l=0}^N \prod_{j=0}^{l-1}\frac{r_j^+}{r_{j+1}^-}}\label{eq:MasterSolution}
\end{equation}
which can be evaluated numerically for any transition rates $r_k^+$ and $r_k^-$. For more details on the exact
solution of the Master equation in the Moran process see for example the work by Claussen and Traulsen \cite{Claussen2005}.

\bibliographystyle{vancouver}

\end{document}